\documentclass[prb,aps,showpacs,twocolumn]{revtex4}
\usepackage{graphicx}

\begin{document}

\title{ Exciton states and tunneling in annealed semimagnetic Cd(Mn,Mg)Te\\
asymmetric double quantum wells}

\author{S.~V.~Zaitsev}
\email{zaitsev@issp.ac.ru}
\author{A.~S.~Brichkin}
\author{P.~S.~Dorozhkin}
\affiliation{Institute of Solid State Physics,RAS,
142432,Chernogolovka, Russia}

\author{Yu.~A.~Tarakanov}
\affiliation{A. F. Ioffe Physico-Technical Institute, Russian
Academy of Science, 194021 St. Petersburg, Russia}

\begin{abstract}

Exciton level structure and interwell relaxation are studied in
Cd(Mn,Mg)Te-based asymmetric double quantum wells (ADQWs) in
normal to plane magnetic fields up to B = 10 T by a steady-state
optical spectroscopy. As grown structures with nonmagnetic CdTe
quantum wells (QWs) were subjected to quick temperature annealing
to introduce Mn and Mg atoms from the barriers inside the wells
resulting in formation of magnetic (MW) and nonmagnetic (NMW) QWs,
respectively, with the concentration of diffused atoms of about
3--5\%. A significant change of exciton energies occurs with
magnetic field: at low fields exciton, localized in the MW, is
higher in energy than that localized in the NMW and efficient
exciton relaxation from the MW to the NMW takes place for all
ADQWs. In all structures the giant Zeeman effect in the MW changes
the energy order of $\sigma^+$-polarized heavy hole (hh) excitons,
localized in different wells, at high B. Levels' crossing is
accompanied by a reverse of tunneling direction without
anticrossing. Calculations of single-particle states and their
change with B indicate that the interwell exciton transfer is
forbidden in the single-particle picture at B $\gtrsim$ 1 T for
all studied structures. Experimentally, nevertheless, a very
efficient interwell relaxation of excitons is found in the whole
magnetic field range regardless of tunneling direction which
evidences about importance of electron-hole Coulomb correlations
in tunneling process. Surprisingly, stress effects are revealed
only in the structure with the narrowest barrier while in ADQWs
with thicker barriers stress relaxes completely. Different
charge-transfer mechanism are analysed in details and elastic
scattering due to strong disorder is suggested as the main
tunneling mechanism of excitons with underlying influence of the
valence band-mixing effects on the hh-exciton transfer in ADQWs
with relaxed stress.

\end{abstract}

\pacs{78.67.-n, 73.40.Gk, 75.75.+a}

\maketitle

\newpage

\section{INTRODUCTION}

During last two decades symmetric and asymmetric double quantum
well (ADQW) nanostructures have attracted high attention due to a
great variety of observed fundamental physical properties as well
as high potential for application in solid state electronics
\cite{Timofeev2004, Butov1,  Butov2, EsakiIEEE}. Among the most
interesting and important phenomena are vertical carrier transfer
- resonant or nonresonant  tunneling and energy relaxation
\cite{Henneberger96, Ferreira92}, energy level hybridization and
state entanglement \cite{DQWth1, Heimbrodt98}. Carrier tunneling
was intensively studied in coupled quantum wells (QWs)
\cite{Ferreira90,Oberly89,Leo90,Kohler91,Kohler92pps,
Henneberger98,Henneberger99,Heimbrodt2003} and in pairs of
correlated quantum dots
\cite{JochenQD2001,MarcusQD2003,ADQDs,ADQDs1}. A possibility to
control of the energy level positions and spatial localization of
carrier states has made ADQW an excellent system for studying of
the tunneling. It was found that excitonic effects are essential
in tunneling process \cite{ShahUltrafast}. Several different
mechanisms were considered to explain a big diversity of effects
observed in tunneling experiments.

The exciton transfer in coupled QWs depends on (i) a
single-particle, electron and hole, level mismatches ($\Delta E_e$
and $\Delta E_h$)  and (ii) an exciton transition energies
mismatch $\Delta E_{ex}$. When both $\Delta E_e$ and $\Delta E_h$
exceed longitudinal optical (LO) phonon energy $\hbar\omega_{LO}$
(21 meV in CdTe)  tunneling of separate single electron and hole
via two LO phonons emission is found to be very efficient
\cite{Bastard89,RosslerLO1}. When such single-particle tunneling
is energetically forbidden for at least of one carrier, but
$\Delta E_{ex} \geq 2 \hbar\omega_{LO}$ , nevertheless efficient
tunneling of the exciton as a whole entity was observed in II-VI
heterostructures \cite{Henneberger96, pss2002}. That was
successfully explained as a two-step process involving indirect
exciton as an intermediate state: strong Coulomb interaction
renormalizes energy spectrum so that transitions forbidden in a
single-particle picture become allowed \cite{Henneberger96}.
Another interesting and theoretically important case takes place
when neither electron nor hole can emit LO phonon, but $\Delta
E_{ex} \geq \hbar\omega_{LO}$. In a single-particle picture,
relaxation of every particle via acoustic phonon (deformation
potential) is expected. This is much slower process with typical
times of hundreds picoseconds \cite{Bastard89, russians}.
Alternative mechanism was theoretically considered by Michl {\it
et al.} \cite{RosslerLO1}, namely a transfer of exciton as a whole
entity through barrier by the emission of a single LO phonon. The
process results  from  admixture of spatially indirect states to
the exciton wave function in ADQW structures via Coulomb
interaction and a strong Fr\"{o}hlich coupling of carriers to the
LO phonons. Exciton tunneling in such situation can dominate over
the separate transfer of electrons and holes if the tunnel
coupling between the wells is strong enough \cite{RosslerLO1}. To
the best of our knowledge, this situation was realized only in a
few works \cite{singleLO,singleLO_B}. Lawrence {\it et al.}
\cite{singleLO} have reported that an efficient tunneling takes
place also at the energy resonance of 1s exciton state of the
narrow QW with a 2s exciton state of the wide QW.

Previously resonances in exciton tunneling were found only in case
of single-particle level resonances for adjacent QWs when using an
external electric field to change mutual levels mismatch : for
electrons \cite{Ferreira90,Oberly89} and for holes
\cite{Leo90,Kohler91,Kohler92pps}. One should emphasize that
careful comparison of the measured and calculated resonant fields
have shown that resonant charge transfer actually involves
excitons tunneling : changes in transfer dynamics are found at
electric field values when direct and indirect transition energies
become equal \cite{Ferreira90,Kohler91,Kohler92SPIE}. These
experimental observations suggest that the resonant tunneling of
exciton-bound carriers is the transfer from spatially direct to
indirect exciton state as was considered in details by Ferreira
{\it et al.} \cite{Ferreira90SSc,Ferreira92}.

Above considerations show a variety of coupling possibilities
realized in ADQW. Besides very strong coupling caused by
Fr\"{o}hlich mechanism in sited papers \cite{Henneberger98,
pss2002, singleLO} it was found that exciton tunneling efficiency
does not drop abruptly but remains still rather strong below one-
or two-LO phonon energy threshold, gradually decreasing and
saturating at small values of $\Delta E_{ex}$. This experimental
findings evidence that other transfer mechanisms are still
efficient below LO phonon threshold. Some other scattering
processes must induce the observed nonresonant carrier tunneling.
For semiconductor heterostructures such mechanisms are
quasielastic scattering through a deformation potential (acoustic
phonons), ionized impurity, interface defect and interface
roughness, or alloy scattering. Tunneling via acoustic phonon
scattering is estimated to be orders of magnitude weaker than the
polar coupling with characteristic tunneling times of about
several hundreds picoseconds or even nanoseconds \cite{Bastard89,
Leo90, russians} and usually should be ruled out. Calculations
have shown that the elastic scattering mechanisms such as ionized
impurity or interface defect assisted interwell transfer are
rather effective and give rise to exciton tunneling times of the
order of several tens picoseconds and, under certain conditions,
even smaller \cite{Bastard89, Ferreira90}.

Besides, strong influence of valence band-mixing effects on heavy
hole (hh) tunneling have been predicted theoretically
\cite{Kohler91r2,Kohler91r3} and observed experimentally
\cite{EsakiAPL,Kohler91r5}. The importance of heavy-light hole
mixing is emphasized in Ref. \cite{Ferreira90} where the authors
argue that the evaluation of transfer time from spatially direct
to indirect exciton states, which involves hole tunneling, cannot
be done without consideration of valence band-mixing effects.
Heavy-light hole mixing influence on exciton tunneling is found to
be effective even at resonant excitation of the hh-exciton
\cite{Kohler91}. Plateau-like dependencies of tunneling time,
observed between sequential resonances as a function of applied
external electric field  \cite{Leo90,Kohler91} indicates
band-mixing effects also \cite{russians}.

One should mention another the most discussed transfer mechanism -
dipole-dipole exciton interaction
\cite{Henneberger98,Takagahara85,Lyo2000} following the pioneering
Takagahara's work \cite{Takagahara85}. Comprehensive experimental
studies \cite{Shah96}, theoretical analysis and numeric
calculations of different exciton transfer mechanisms in ADQWs
including dipole-dipole interaction \cite{Lyo2000} have given too
small transfer rates ($<$ 10 $^9$ sec$^{-1}$), unable to explain
experimentally observed values \cite{Shah96}. On the other side,
in elegant work by Kim {\it et al.} \cite{Kim96,KimSSC}, high
tunneling transparency was demonstrated experimentally for thick
ternary AlGaAs barriers while opposite situation takes place for
equivalent  GaAs/AlAs digital barrier. These investigations
clearly pointed to often forgotten fact that usually barrier
material is an alloy semiconductor. If the alloy is inclined to
clustering like GaAs in AlGaAs barriers one can expect percolation
in charge transfer through low potential channels in the barrier.
Simple model considered by Kim {\it et al.} \cite{Kim96,KimSSC}
confirms the conclusion.

This short review shows the dominant role of excitons in tunneling
process in ADQWs but the transfer mechanism strongly depends on
particular parameters and/or structure design. Nevertheless, in
some experimental conditions the exact nature of the process is
unclear and still under discussion.

In this paper we report the results of experimental studies of
exciton level structure and its influence on charge transfer in
semimagnetic CdTe/Cd(Mg,Mn)Te-based ADQWs as a function of
magnetic fields and barrier width by using steady-state
photoluminescence (PL) and PL excitation (PLE) spectroscopy. Great
advantage of semimagnetic or, in another words, diluted magnetic
semiconductors (DMS) is that a giant Zeeman effect makes possible
continuous tuning of a band-gap E$_g$ and exciton energies by
external magnetic field due to strong s,p-d exchange interaction
between free band carriers and localized d-states of magnetic ions
\cite{Furdyna}. External magnetic field allows to vary the
interwell coupling in the DMS ADQW after preparation, when the
barrier width is fixed. We have found very efficient hh-exciton
interwell relaxation in investigated ADQWs. Possible
exciton-transfer mechanisms are analysed in details.

\section{EXPERIMENTAL DETAILS}

Cd$_{0.8}$Mg$_{0.2}$Te/CdTe/Cd$_{0.8}$Mg$_{0.2}$Te/CdTe/Cd$_{0.8}$Mn$_{0.2}$Te
undoped ADQWs were grown by molecular beam epitaxy on
thick CdTe buffer deposed on (001)-oriented CdZnTe substrate
(Fig.~1).  As-grown structures have two nonmagnetic pure 6-nm wide
CdTe QWs separated by Cd$_{0.8}$Mg$_{0.2}$Te barrier with L$_B$=
3, 4 and 6 nm.

\begin{figure}%h
\begin{center}
\includegraphics[width=8.5cm]{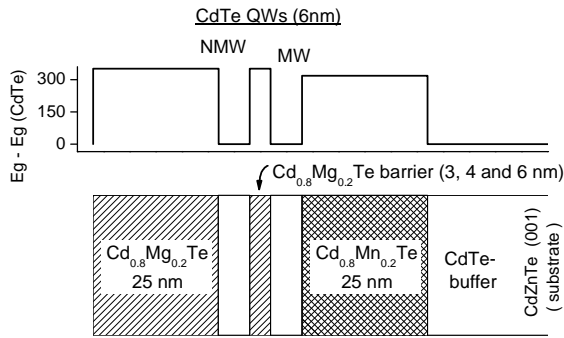}
\end{center}
\caption{Outline and band gap alignment of as grown ADQW
structures. } \label{spectra}
\end{figure}

Effective thermal inter-diffusion between the barriers and CdTe
QWs is promoted by use of SiO$_2$ upper mask and subsequent rapid
temperature annealing (RTA) once (L$_B$= 3 and 6 nm) or two times
(L$_B$= 4 nm) at 400 $^o$C for one minute to promote diffusion of
Mn and Mg atoms from the barriers into the QWs. 80 nm thick
SiO$_2$ mask was deposed by electron beam lithography and lift-off
after procedure. As we reported previously, diffusion is strongly
enhanced below the SiO$_2$ mask as compared to non-covered areas
\cite{RTAmy} and results in the increase of band-gap below the
masked areas, inducing a lateral confinement potential up to 0.3
eV \cite{RTAmy} depending on the sample and processing parameters.
It was shown that RTA technique allows to vary the QW energy gap
in a controllable way with a good optical quality \cite{RTA1,
RTA2,RTAmy}. QW, located between two CdMgTe barriers, incorporates
only Mg atoms after RTA and is referred to as nonmagnetic well
(NMW) whereas that with CdMnTe barrier contains both Mg and Mn
atoms and is referred to as magnetic well (MW).

Measurements  were done in superfluid He (bath temperature
$T\simeq 1.8$ K) in a cryostat with superconducting magnet in
Faraday geometry. Dye-laser (Pyridin1 dye), pumped by continuous
Ar$^+$-ion laser, was used for excitation. Circularly polarized
($\sigma^{\pm}$) laser beam and photoluminescence (PL) signal were
formed with quarter-wave plates and linear polarizers. Sample
surface was covered with opaque metallic (Au) mask leaving open
small holes of 50 $\mu$m size. PL spectra were detected with a
double-stage 0.82 m monochromator and CCD camera. PLE spectra were
recorded as a spectrally integrated signal at low-energy spectral
tail of PL band versus excitation energy. Such an experimental
setup was applied to minimize a stray light when recording very
broad spectral bands usual for nanostructures based on ternary
semiconductor materials \cite{revDimaYakovlev}.

Maximum dye-laser excitation power density was $\simeq$ 10
W/cm$^2$ to keep reasonable signal to noise ratio and to avoid
overheating which destroys sp-d interaction in DMS semiconductors
\cite{Furdyna}. Temperature of magnetic Mn-ions spin system in the
excitation spot, estimated from magnetic field dependencies of
exciton energies, was about 5 K.

\section{EXPERIMENTAL RESULTS}

Figure 2 displays polarized magneto-PL spectra of ADQWs. Emission
was excited with $\sigma^-$ polarized laser at $\sim$ 100 meV
above exciton transitions but below the barrier band gap. The
spectra of all structures display broad bands with a halfwidth of
6--8 meV, characteristic for ternary II-VI materials. At low B
spectral positions and line intensities are nearly B-independent
in all samples, as it usually does in the NMW. At higher B, at
some particular field B$_C \approx$ 6, 3 and 2.5 T in samples with
L$_B$= 3, 4 and 6 nm correspondingly, a red shift of the
$\sigma^+$ component increases strongly while its intensity
increases gradually, which is characteristic to excitons in the
DMS QWs. The higher energy $\sigma^-$ component, in contrast,
displays a very weak shift characteristic to the NMW exciton in
the whole magnetic field range. Its intensity decreases with B in
all samples, the stronger decrease for ADQW with bigger L$_B$.

\begin{figure}%h
\begin{center}
\includegraphics[width=8.5cm]{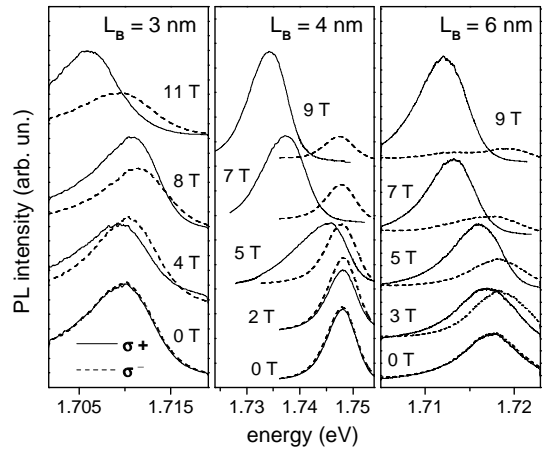}
\end{center}
\caption{ Magneto-PL spectra of ADQWs with $L_B=$ 3, 4 and 6 nm in
Faraday geometry and intra-well $\sigma^-$-polarized excitation at
$\sim$ 100 meV above exciton transitions. } \label{spectra}
\end{figure}

Excited exciton states were studied by the PLE. Fig.~3 represents
polarized PLE spectra recorded at different magnetic fields in
both polarizations. The lowest hh-exciton transition is
B-independent at low fields in all structures. Thereby we ascribe
it to the spatially direct transition in the NMW and label as hh
(NMW) (Fig.~3). The first excited hh-exciton, labelled as hh (MW),
behaves similar to that in DMS heterostructures
\cite{revDimaYakovlev}: in $\sigma^+$ excitation polarization its
energy decreases quickly with magnetic field while in $\sigma^-$
polarization - increases with a giant Zeeman splitting up to 60
meV. Thereby it is ascribed to the direct transition in the MW. At
some magnetic field, close to B$_C$, hh-exciton in the MW crosses
that in the NMW. The higher exciton transitions are associated
with a light hole (lh) in the MW and NMW QWs. No evidence of
Landau level formation is found till B = 12 T as expected for QWs
with strong potential fluctuations.

\begin{figure}%h
\begin{center}
\includegraphics[width=8.5cm]{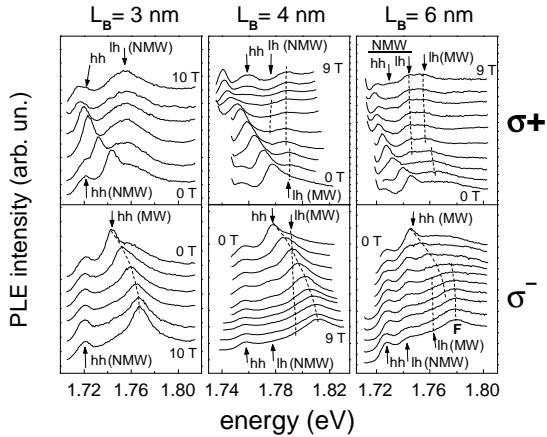}
\end{center}
\caption{ Polarized PLE spectra recorded at $\sigma^+$-polarized
excitation (upper panels) and $\sigma^-$-polarized excitation
(lower panels). Magnetic field increases from the bottom to the
top for $\sigma^+$ polarization and vice versa for $\sigma^-$
polarization with a step in 2 T for the ADQW with L$_B$ = 3 nm and
1 T for ADQWs with L$_B$ = 4 and 6 nm. Dashed lines are to guide
the eye. } \label{spectra}
\end{figure}

Magnetic field dependencies of the exciton transition energies in
both circular polarization are summarized in Fig.~4. Energies for
hh and lh-exciton transitions are estimated with the accuracy of
about 1 meV and 2 meV, respectively, because of the broad
inhomogeneous spectral width of exciton bands. Nevertheless,
systematic change of exciton energies with magnetic field can be
determined and comparison with calculated dependencies can be
made.

\begin{figure}%h
\begin{center}
\includegraphics[width=8.5cm]{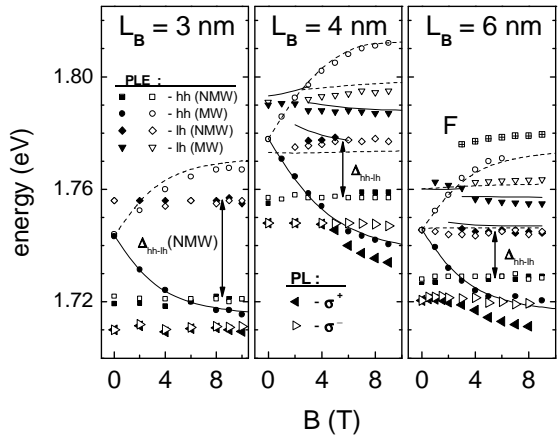}
\end{center}
\caption{ Magnetic field dependencies of the exciton transition
energies obtained from PLE (small symbols) and PL spectra (large
symbols). Filled symbols and solid curves - for $\sigma^+$
polarization, open symbols and broken curves - for $\sigma^-$
polarization; curves - are calculated dependencies acc. to the
model (see text). $\Delta_{hh-lh}$ depicts hh-lh splitting in the
NMW. } \label{spectra}
\end{figure}

Spatially indirect hh-excitons have not been observed both in PL
and PLE spectra even in the structure with smallest L$_B$. On the
other hand, it is known that the energy of indirect exciton in DMS
ADQW is very sensitive to the structure parameters \cite{Sugakov}.
On the whole PL and PLE data for the ground exciton transitions
agree with each other with the values of stokes shift of 7--9 meV.

\section{DISCUSSION}

\subsection{ General remarks}

Spectral band, assigned to the hh-exciton of MW, splits in
magnetic field into two components with opposite, $\sigma^+$ and
$\sigma^-$, polarizations corresponding to J = +1 and J = --1
states of bright excitons \cite{IvchenkoBook}. J = +1 and J = --1
excitons in PLE spectra demonstrate the giant Zeeman effect in the
MW well up to 60 meV  with a saturation at high magnetic fields
(Fig.~4). Thus, a diffusion process during RTA treatment gives
rise to a significant concentration of magnetic Mn atoms in the
MW. The main contribution in splitting is provided by the strong
sp-d exchange interaction of electrons and holes in semimagnetic
CdMnTe QWs with localized Mn$^{2+}$ magnetic moments
\cite{revDimaYakovlev}. As a result, the giant Zeeman splitting of
$\sigma^+$ and $\sigma^-$ polarized excitons in DMS semiconductors
up to hundreds meV takes place with a huge negative $g_{hh}$ and
positive $g_e$ effective g-factors of the heavy hole and electron,
respectively \cite{Furdyna}.

Ground exciton transition in the MW at B $>$ 0 is $\sigma^+$
polarized \cite{revDimaYakovlev} and corresponds to optical
transition between the upmost state with $J_z=-3/2$ moment
projection in the valence band $\Gamma_8$ (heavy hole state) and
the lowest state with $S_z=-1/2$ spin projection in the
conductivity band $\Gamma_6$. Very quick, till a picosecond time
scale, carriers relaxation from the excited states, spin-split  at
finite B, to the ground one is induced by the strong sp-d exchange
interactions between free carriers and Mn$^{2+}$ ions
\cite{SemenovDyakonov2001, CrookerAwschalom97} and usually gives
rise to observation in PL of only $\sigma^+$ polarized exciton
transition at B $\gtrsim$  1 T both in 3D and 2D situation.

As a test of the used PLE methodic we recorded PLE spectra at
different spectrometer positions and have not found noticeable
difference in spectra shape except of intensity. According to the
Ref. \cite{longRlocaliz} this finding points to the absence of
long-range localizing potential while very broad exciton bands
(6--8 meV) evidence for a short-range disorder due to interface
and/or alloy fluctuations, naturally expected in the investigated
ADQWs after RTA treatment.

Strong Zeeman shift, observed for the MW exciton transition,
evidences about high effectiveness of the interdiffusion between
the barriers and CdTe QWs, promoted by the RTA and enhanced by
using of the SiO$_2$ upper mask. Good optical quality as compare
to the other as-grown II-VI ternary QW heterostructures
\cite{Henneberger96,pss2002} shows up perspectives of this
technology for controllable engineering of optical properties in
nanostructures. For our purposes the possibility to vary the
interwell coupling and mutual exciton levels in such ADQWs by the
external magnetic field, when the structure parameters are fixed
after preparation, allows to  study interwell exciton tunneling.

\subsection{Interwell exciton relaxation}

At low magnetic fields (B $<$ B$_C$)  most of the exciton
recombination is observed at the lowest in energy NMW exciton
transition at nonresonant excitation (Fig.~2). PLE data
demonstrate that at resonant excitation of the MW hh-exciton an
effective interwell exciton transfer takes place since the MW and
NMW hh-exciton bands have comparable intensities in the PLE
spectra (Fig.~3). This finding evidences that the interwell
relaxation time of excitons, excited in the MW, is markedly
smaller than their recombination time and during lifetime they
mostly relax to the NMW. Above the crossing field B$_C$, a
transfer in the opposite direction also takes place due to the
same reason (PLE signal is detected at the tale of the MW PL
band). Though hh-exciton, attributed to the NMW, is less
pronounced at B $>$ B$_C$ in PLE spectra with $\sigma^+$ polarized
excitation because of strong spectral bands overlap and
broadening, it is clearly seen in $\sigma^-$ excited  PLE spectra
(Fig.~3). Exciton levels crossing at B$_C$ in $\sigma^+$
polarization  is accompanied by a reverse of tunneling direction
without spectral peculiarities in the crossing point. Absence of
anticrossing behaviour points to the absence of hh-exciton level
interaction and, thus, to an incoherent nature of exciton
tunneling \cite{ShahUltrafast}.

\subsection{Band alignment and energy levels calculations}

To understand behavior of exciton transitions and interwell
exciton tunneling  in magnetic field we have performed numeric
calculations of single-particle energies in $\Gamma$ point (zero
in-pane wavevector) in the investigated ADQWs. Wave functions and
level energies were obtained by solving of Schrodinger equation
with appropriate boundary conditions. Band gap dependencies of
E$_g$ of Cd$_{1-x}$Mg$_{x}$Te and Cd$_{1-x}$Mn$_{x}$Te on the Mg
and Mn concentrations are known from Ref. \cite{revDimaYakovlev},
while the change of E$_g$ with magnetic field in the DMS barrier
and in the MW are described by a modified Brillouin function with
accurately tabulated parameters of the sp-d exchange interactions
in Cd$_{1-x}$Mn$_{x}$Te \cite{Furdyna,revDimaYakovlev}. Effective
electron, heavy- and light-hole masses are taken from Ref.
\cite{bandOffsetWaag}. Conduction-to-valence band offset for
Cd(Mn,Mg)Te heterostructures has been shown to obey usual for
II-VI compounds 2:1 ratio rule \cite{bandOffsetWaag}. In our
calculations the potential profiles of conductivity and valence
bands were approximated by the rectangular shape. Though this
approximation is rather simple because real concentration profiles
of Mn and Mg after RTA treatment depend on the growth conditions
and post-grown treatment \cite{RTAbandErr,IFmixin1,IFmixin2}, it
allows  to describe well the experimental data as will be shown
below. Stress induced band shifts are used as adjustable
parameters. We also neglect a small Zeeman splitting in the NMW
($\approx$ 1 meV at B = 10 T) \cite{gFactorDimaYak}  - markedly
smaller than the PLE bands spectral linewidth.

Concentrations of diffused Mn in the MW ($x_{Mn}$) and Mg atoms in
the NMW ($x_{Mg}$) are  evaluated from the shifts of exciton
transition energies above the pure CdTe  QW states in both wells
at B = 0 T and, also, from a fit of magnetic field dependencies of
hh-exciton transition in the MW. They are found to be: $x_{Mn}
\approx $ 3.0, 5.1 and 3.1 and \% $x_{Mg} \approx $ 4.1, 5.2 and
4.4 \% for samples with barrier 3, 4 and 6 nm correspondingly.

Excitonic effects such as binding energy and oscillator strength
are known to be strongly enhanced in II-VI heterostructures
\cite{IvchenkoBook} due to confinement. Modification of ADQW
potential profile with magnetic field and successive change of
confinement must influence on the exciton intensities and binding
energies. In calculations of the exciton binding energies we have
used general formulas derived in Ref. \cite{ExRydberg} and valid
for an arbitrary confinement potential, which are shown to give
good accuracy (less than 1 meV) also in the case of strongly
coupled ADQWs. This approach is exact only at zero magnetic field,
systematically underestimating exciton binding energy with
increasing B, especially  at high fields, when magnetic length
$l_B$ is comparable with the in-plane exciton radius
\cite{IvchenkoBook}. Nevertheless, because the exciton in-plane
radius is $\approx $ 7 nm in all ADQWs, according to variational
estimations basing on the Ref. \cite{ExRydberg}, and becomes
comparable with $l_B$ only at B $\approx $ 13 T, used approach
allows  to track qualitatively  and even semi-quantitatively
changing of the exciton binding energy with B in the field
dependent potential.

\subsection{Results of calculations}

Calculated potential profiles of $S_z=-1/2$ conductivity and $J_z
= -3/2$ valence bands, wave functions  and single-particle energy
levels, corresponding to $\sigma^+$ polarization of optical
transition, are presented in Fig.~5 for the ADQW with L$_B$= 4 nm
at three values of magnetic field: B = 0, 2 and 6 T (below and
above B$_C$). At zero field both electron and hh ground states,
localized in the MW, are higher in energy than those in the NMW.
The giant Zeeman effect in the MW and DMS barrier dramatically
decreases band energies with the main change in the valence band
as the exchange interaction constant is four times stronger for hh
states than that for the electron in the CdMnTe \cite{Furdyna}.
The hh levels of the DMS and NMWs cross each other at some
specific field value B$_H\approx$ 1 T.

\begin{figure}%h
\begin{center}
\includegraphics[width=8.5cm]{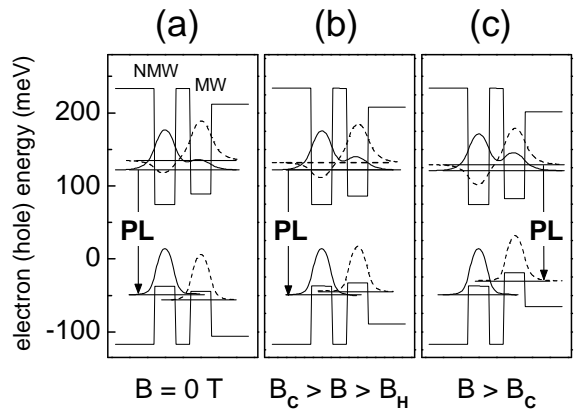}
\end{center}
\caption{ Calculated electron and hole band potential profiles for
$S_z=-1/2$ and $J_z=-3/2$ states, corresponding to
$\sigma^+$-polarized optical transitions, for ADQW with L$_B$ = 4
nm at B = 0 (a), 2 (b) and 6 T (c). Electron and hole wave
functions in the NM QW (solid lines) and NMW QW (dashed lines) are
shifted according to state energy. Arrows indicate observed PL
transitions. } \label{spectra}
\end{figure}

Calculated magnetic field dependencies of the single-particle
ground and first excited state energies  for electron (e1,e2),
heavy (hh1,hh2) and light hole (lh1,lh2) in both spin states are
presented in Fig.~6 for ADQWs with L$_B$= 4 and 6 nm. The energies
are given relatively to the conductivity and valence band edges of
bulk CdTe. These particular ADQWs are chosen for detailed analysis
and comparison with the experiment because they are supposed to be
unstressed as will be discussed below. For the ADQWs with L$_B$ =
3 nm the dependencies of the electron and hh levels are
qualitatively  same as in another structures with the exception of
the lh levels which are split much more (Fig.~4).

\begin{figure}%h
\begin{center}
\includegraphics[width=8.5cm]{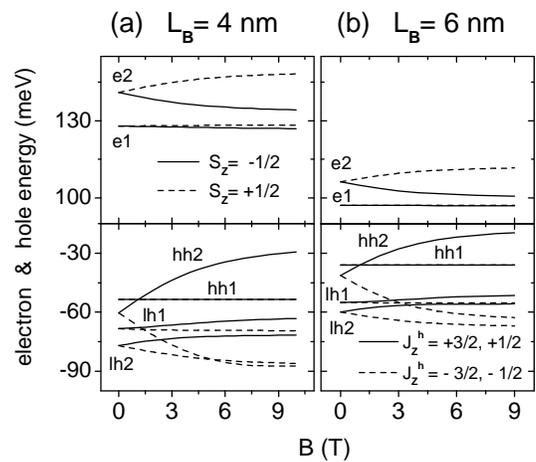}
\end{center}
\caption{ Calculated magnetic field dependencies of
single-particle ground and 1st excited state energies (relatively
to the band edges of pure CdTe) for electrons (e1,e2), heavy
(hh1,hh2) and light holes (lh1,lh2) for ADQWs with L$_B$= 4 (a)
and 6 nm (b). Spin states are marked in figure. Solid lines
correspond to states for $\sigma^+$-polarized optical transitions,
broken lines - for $\sigma^-$. } \label{spectra}
\end{figure}

Figure~6 shows that hh levels of the MW and NMW  with $J_z=-3/2$
reverse their sequence with magnetic field  due to the giant
Zeeman effect in the MW at B$_H\approx$ 1.0 T in both ADQWs. No
levels crossing is found for the electrons: the ground electron
level e1, mainly localized in the NMW, has lower energy than level
e2, localized in the MW, in the whole magnetic field range
(Fig.~6).

Figure~7(a) presents integrated squares of corresponding wave
functions $\Psi_e(z)$ and $\Psi_h(z)$ (probabilities) for
electrons and hh (left panel) and lh states (right panel) in the
regions of the NMW and MW for the ADQW with L$_B$ = 4 nm. Hole
spin representation is used below - the hole "spin" projection is
opposite to the moment projection of the state in the valence band
$\Gamma_8$. Calculations show that heavy holes are effectively
decoupled in adjacent wells: probability $P_{~leak}$ not to be in
its "own" well is less than 10$^{-4}$ except of a very narrow
region of B near B$_H$. Moreover, hh states are strongly localized
even in the ADQW with the narrowest barrier L$_B$= 3 nm:
$P_{~leak}<$ 10$^{-3}$. Contrary, electrons and lh states with
$J_z^h= +1/2$ demonstrate strong coupling. Besides, with
increasing B these light holes change their location to the
opposite QWs gradually, in a very wide magnetic field range 1 -- 8
T (Fig.~7(a), right panel), following band gap shrinkage in the
MW, but still having a significant probability to be in the
opposite well: $P_{~leak}> 4 \%$ at high fields. Light holes with
opposite spin $J_z^h=$ -1/2 preserve original, same to B = 0 T,
localization. Such behavior reflects delocalized nature of light
holes in investigated ADQWs due to much smaller value of a lh mass
in the quantization direction \cite{bandOffsetWaag} and smaller
barrier value for the lh states.

\begin{figure}%h
\begin{center}
\includegraphics[width=8.5cm]{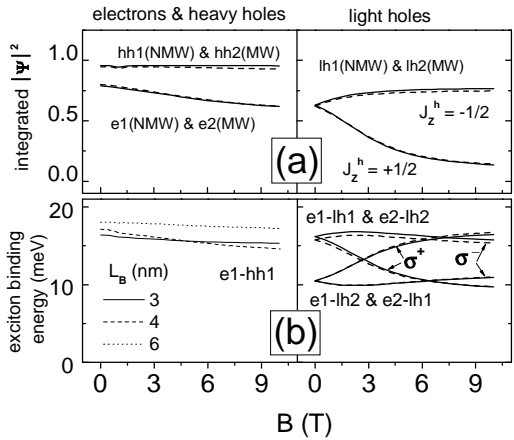}
\end{center}
\caption{ (a) Integrated squares of wave functions $\Psi_e(z)$ and
$\Psi_h(z)$ vs B for electrons and hh (left panel) and lh states
(right panel) in the regions of the NMW and MW (marked in figure)
for ADQW with L$_B$ = 4 nm. (b) Exciton binding energies for
$\sigma^+$-polarized hh-excitons (left panel, all ADQWs) and
lh-excitons in the ADQW with L$_B$ = 4 nm (right panel). e1-lh1
and e1-lh2 (e2-lh2 and e2-lh1) excitons are shown by solid
(dashed) lines, respectively. $\sigma^+$ ($\sigma^-$) exciton
polarizations are marked in figure. } \label{spectra}
\end{figure}

Exciton binding energies as a function of magnetic field are shown
in Fig.~7(b) for $\sigma^+$ polarized hh-excitons for all ADQWs
(left panel) and for lh-excitons in the ADQW with L$_B$= 4 nm
(right panel). Interestingly, it turned out that in each ADQW
binding energies of hh  exciton in the MW and NMW have very close
values in the whole magnetic field range (the difference is less
than 0.5 meV) which is caused by strong hh confinement. One can
see a strong influence of the the spatial localization on the
binding energy: first, hh-excitons in ADQW with wider barrier have
higher binding energy value due to stronger electron localization.
Secondly, binding energy of the lh-excitons correlates with the
localization degree of the light holes: the smaller overlap of lh
wavefunction with electron - the lower corresponding binding
energy and vise versa.

According to calculations for ADQW with L$_B$= 4 and 6 nm, NMW
lh-exciton in $\sigma^+$ polarization is ascribed to e1-lh1
transition at low magnetic fields (B $<$ 2 T) when both carriers
are localized mainly in the NMW (Fig.~7(a)). With increasing field
lh1 continuously delocalizes and changes its location to the MW.
At higher fields (B $>$ 5 T) e1-lh2 transition can be considered
as the NMW lh-exciton because in this field range lh2 level
changes its localization to the NMW while e1 state is always
mostly localized in the NMW (and e2 - in the MW). In $\sigma^-$
polarization, like zero field situation, e1-lh1 transition is
being the NMW lh-exciton because for these carrier spin states
band profile doesn't change relative QWs alignment, preserving
carrier localization.

Similarly, the identification of lh-exciton can be done in the MW.
For this exciton transition lh1 and lh2 levels interaction is seen
most prominantly. According to calculations, MW lh-exciton at low
B starts as e2-h2 transition, when both carriers are localized
mainly in the MW, and have the highest transition energy. At
higher fields (B$>$ 3 T) MW exciton in $\sigma^+$ polarization
quickly decreases its energy and can be assigned to e2-lh1
transition in accordance with the change of lh localization. In
$\sigma^-$ polarization the MW lh-exciton is ascribed to e2-lh2
transition at all B because both carriers are localized in the MW.

Calculated lh-exciton transition  energies for discussed ADQWs
with L$_B$= 4 and 6 nm  are depicted  together with experimental
ones in Fig.~4. Data for the lh-excitons are presented in the
field regions where they have significant intensities according to
the above discussion. One can see rather satisfactory agreement
between calculated values and measured ones with worse agreement
for the lh-excitons. Is not surprising for the hh-excitons because
model parameters were chosen to fit their experimentally found
energies. As to the lh-excitons, comparison with the experiment is
complicated due to broad spectral width of exciton bands in PLE
and smaller value of lh microscopic transition matrix elements as
compare to hh transitions \cite{IvchenkoBook}. Four possible
combinations of light hole excitons in every polarization (two
electrons times two lh holes) are reduced to only two strong
transition at low and high magnetic fields due to definite
localization of the light holes thus complicating analysis only at
middle B values. Analysis can be entangled also by interplay
between intrawell and interwell exciton relaxation kinetics and
reversing of exciton transfer direction between QWs at B$_C$. As
an example of such an ambiguity, in the ADQW with L$_B$= 4 nm it
is difficult to identify the NMW lh-exciton transition in
$\sigma^+$ polarization whereas it is clearly seen in $\sigma^-$
polarization. Contrary, in the ADQW with L$_B$ = 6 nm lh-exciton
from the NMW is clearly seen in both polarizations.

Calculations for the ADQW with L$_B$= 3 nm  have shown that
e1(NMW)-hh2(MW) indirect exciton is the ground energy state at
high B. Experimentally, nevertheless, indirect exciton is not
observed in this structure as well as in another ones. On the
other hand, it is known that the energy of indirect exciton in DMS
ADQWs is very sensitive to the structure parameters \cite{Sugakov}
and we ascribe this contradiction to the weakness of calculations
for the ADQW with the narrowest barrier.

In addition to identified lh-exciton transitions Fig.~3(c)
displays  $\sigma^-$ polarized transition arising at B$>$ 4 T for
the ADQW with L$_B$= 6 nm (noted as F). It cannot be assigned to
transitions with participation of higher electron or hole levels
because they have much higher energies. We suppose that it appears
due to hh-lh valence band-mixing between hh2 ($J_z^h=$ -3/2) and
lh2 ($J_z^h=$ -1/2) levels, localized  in the MW and converging at
high fields (Fig.~6). Another evidence for such a mixing is
anomalously broad MW-assigned hh-bahd in the ADQW with L$_B$ = 4nm
(Fig.~3(b)) in which, according to calculations, hh2 and lh2
levels cross at high fields and stay very close (less than 3 meV
at B$>$5 T). Both hole states are localized in the MW, hh2 -
completely and lh2 - with a probability $>82 \%$ and can be
analyzed as first heavy and light hole states in a single QW.
Symmetry analysis presented in Ref. \cite{BauerAndo88} (table II
therein) shows that there is only 2p$^-$lh-exciton state with the
same symmetry $\Gamma_{7g}$ as a ground 1shh-exciton, which can
mix with 1shh in $\sigma^-$ polarization. Symmetry requires also
that 2p$^-$lh-exciton should have odd single-particle wave
functions \cite{BauerAndo88} (negative parity of the product
$\Psi_e(z) \ast \Psi_h(z)$) and only violation of mirror symmetry
in the MW can give rise to such considerable strength of optically
forbidden exciton state. Another $\Gamma_{7g}$ state, 3d$^\pm$lh,
has orbital moment projections $m = -3, \pm2$ and +1 and  doesn't
couple to the $\sigma^-$-polarized optical field. We suppose it is
a violated symmetry in ADQWs  after RTA treatment
\cite{IFmixin1,IFmixin2,MnSegreg} due to two times bigger
diffusion coefficient of Mn atoms than that of Mg in CdTe matrix
\cite{RTAcrystGrowsTonnies} which is responsible for the origin of
this band in the MW.

Thus, calculations in the simplest model with rectangular shape
potential profile describe very well hh-exciton magnetic field
behavior in the studied ADQWs. As for the lh-excitons,
correspondence is worse due to smaller $m_{lh}^z$ mass and higher
lh energies, inquiring knowledge of exact band potential shape.
Basing on our calculations, we conclude that the heavy hole levels
are completely decoupled, while light hole levels interact
strongly with pronounced coupling.

\subsection{Stress effects in annealed ADQWs}

It is known that in heterostructures with a single QW stress is
fully accommodated by the QW with the barriers being unstressed
completely \cite{StressReview}. In calculations of light hole
energies for ADQWs with L$_B$= 4 and 6 nm the stress induced
additional shift of lh levels is set to zero. Only in this case
calculations provide satisfactorily description of exciton
transitions considered above and, particularly, explain small
values of hh-lh splitting $\Delta_{hh-lh}$ for these ADQWs
(Fig.~4). Stress effects in the QWs are completely neglected only
for these two particular structures except of one with L$_B$= 3
nm, where lh-exciton in the NMW is sufficiently higher split in
energy: $\Delta$E$_{hh-lh}\approx$ 35 meV. This value is much
bigger than one expected without stress. Such a big splitting is a
result of considerable stress inside QWs in the ADQW with the
narrowest barrier.

Stress relaxation effects in the annealed ADQWs with L$_B$= 4 and
6 nm as a result of RTA treatment are in accordance with the
results of thorough experimental investigations and numerical
simulation of diffusion process in asymmetric CdMnTe/CdTe/CdMgTe
quantum wells \cite{RTAcrystGrowsKossut}. In this work authors
indicate that full stress relaxation during the RTA procedure is
significant to explain their and reported by T\"{o}nnies {\it et
al.} \cite{RTAcrystGrowsTonnies} results. To understand the
experimentally observed difference of stress relaxation in ADQWs
with  different barriers one should suggest: (i) the stress can
partly relax in the interwell barrier region, i.e., to be adopted
by the barrier; (ii) the strongest relaxation occurs at some
optimal barrier width comparable with the QW width, because
relaxation effect should vanish in utmost cases for very narrow
(L$_B$= 3 nm) as well as for very wide barriers. We don't have
direct references concerning this phenomenon except of Ref.
\cite{RTAcrystGrowsKossut} and make this speculation as a
suggestion to explain the experimental data. Besides, ADQW with
L$_B$= 4 nm is treated twice at elevated temperature thus the
stress relaxation effect for this structure should be stronger.

\subsection{Exciton tunneling mechanisms in annealed CdTe-based
ADQWs}

\subsubsection{Semiclassical single-particle approach}

In the single-particle picture,  tunneling of the heavy hole  with
zero in-plane hh kinetic energy from the MW to the NMW is
prohibited at B$>$B$_H$ since the final hole state is higher in
energy. On the other hand, exciton transition in the MW has higher
energy than that in the NMW until B$_C$ and, as  PLE experiment
definitely shows, effective tunneling of the hh-excitons from the
MW to the NMW at B $<$ B$_C $ takes place. Vice versa, at B $>$
B$_C $, effective hh-exciton tunneling from the NMW to the MW
occurs as discussed at the beginning of this section. Electron
level e1, localized in the NMW, has lower energy than that in the
MW (e2) in the whole magnetic field range (Fig.~6) which prevents
tunneling of the electron with zero in-plane kinetic energy from
the NMW to the MW at B $>$ B$_C$, again in contradiction with the
experiment. Thus, in contrast with the single-particle picture,
experiment indicates very efficient exciton transfer in the field
range B$>$B$_H$, prohibited for the tunneling of uncorrelated free
carriers. This contradiction emphasizes the importance of exciton
correlation in charge transfer processes in semiconductor
heterostructures and shows that the tunneling direction is
governed only by the exciton transition energies in the adjacent
QWs in accordance with the results of Refs.
\cite{Henneberger96,Henneberger98,pss2002,singleLO}.

In discussing the nature of exciton transfer mechanism in
investigated ADQWs, one has to mention that the semiclassical,
single-particle estimation of the heavy hole tunneling time
\cite{tau_semiclass} gives values $\approx$ 8 ps, 65 ps and 2.7 ns
for ADQWs with barriers L$_B$= 3, 4 and 6 nm respectively. On the
other hand, free exciton radiative lifetime in
CdTe/Cd(Mg,Mn)Te-based QWs falls in range of 80--150 ps depending
on the QW parameters \cite{singleLO,tauCdTe,tauSpinCdTe}. Thus,
only for ADQWs with L$_B$= 3 and 4 nm estimated tunneling times
are reasonable to explain experimentally observed efficient
exciton transfer during exciton lifetime, being completely
unappropriate for the structure with the widest barrier of 6 nm.

\begin{figure}%h
\begin{center}
\includegraphics[width=8.5cm]{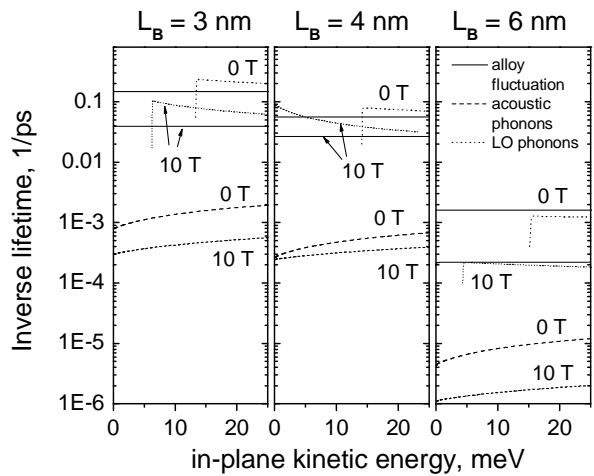}
\end{center}
\caption{ Calculated dependencies of the scattering rates vs
in-plane hh kinetic energy between hh2 and hh1 states via alloy
fluctuations (solid lines), emission of the acoustic phonons
(dashed lines) and LO phonons (dotted lines) at B = 0 (thick
lines) and 10 T (thin lines). } \label{spectra}
\end{figure}

To clarify the question about  tunneling times in the studied
ADQWs, exact calculations of the hh interwell tunneling rates
(inverse tunneling times) have been done in a single-particle relaxation
picture. We considered only the most reasonable in the studied
undoped structures transfer mechanisms such as relaxation via the
emission of the acoustic phonons \cite{Bastard89}, LO phonons
\cite{phononScatt} and alloy fluctuation elastic scattering
\cite{BasuAlloy}. For the alloy fluctuation elastic scattering we
used the value of scattering potential of GaAs because of the
absence of this value for Cd(Mn,Mg)Te compounds in the literature.
Calculated dependencies of the scattering rates on the initial
in-plane hh kinetic energy from the higher state hh2 to the ground
state hh1, which are localized in different QWs, are depicted in
Fig.~8 for B = 0 and 10 T. At B = 0 T hh tunneling takes place
from the MW to the NMW while at B = 10 T - contrariwise. Figure~8
shows that the relaxation via acoustic phonons is much slower
process with typical times of nanoseconds and thus can be
neglected. One can see that the elastic alloy-disorder relaxation
mechanism does't depend on the in-plane hh kinetic energy
\cite{BasuAlloy} and do provide transfer rates sufficient for
effective interwell hh tunneling in ADQWs with L$_B$= 3 and 4 nm:
calculated transfer times are of about several tens picoseconds
(or even smaller for the ADQW with L$_B$= 3 nm) do not exceed typical
exciton recombination times ($\sim$~100 ps in CdTe QWs
\cite{tauCdTe}). Usually the strongest scattering mechanism
through LO phonons in the studied structures has an energy
threshold except for high B in the ADQW with L$_B$= 4 nm
(Fig.~8(b)). According to the single-particle calculations,
LO-threshold for energy difference between the hh1 and hh2 states
in this structure vanishes at B $>$ 7.5 T (Fig.~6(a)) which
explains evident growth of the NMW hh-exciton band in the PLE
spectra (Fig.~3, L$_B$= 4 nm). As to the ADQW with L$_B$= 6 nm,
neither of the considered mechanisms can provide sufficient
interwell transfer rates in agreement with simple semiclassical
estimations. Thus, very weak tunnel coupling between the hh states
for the widest ADQW gives rice to very low interwell transfer
rates, which cannot explain efficient hh-excitons transfer,
experimentally found in the presented studies.

\subsubsection{Barrier quality}

In our previous publications of the influence of RTA treatment on
the optical properties of CdTe/(Cd,Mg)Te annealed nanostructures
were investigated \cite{RTAaplQWs,RTAapl94Tonnies}. We have not
found any sign of (i) segregation phenomena in this material
system and (ii) deviation of Mg diffusion from the classical
linear Fick's law. Same conclusion is made in the studies of
nonmagnetic CdTe/CdMgTe QWs \cite{MgDiffQWs}. Moreover,
comparative investigations of diffusion process in RTA procedure
have revealed same activation energy and comparable values of Mg
and Mn atoms diffusion coefficient in CdTe matrix
\cite{RTAcrystGrowsTonnies}. In magneto-optical studies in the DMS
CdTe/CdMnTe heterostructures no segregation effect of Mn was found
also
\cite{IFmixin1,IFmixin2,RTAcrystGrowsTonnies,MnSegreg,RTAcrystGrowsKossut}.

Thus, we have no hints to suppose the existence of low potential
conductivity channels in the barrier, caused by alloy segregation
or clustering in the investigated material system Cd(Mn,Mg)Te,
similar to reported by Kim {\it et al.} for GaAs/AlGaAs
heterostructures \cite{Kim96,KimSSC}.

\subsubsection{LO-phonon exciton tunneling}

We have analyzed also another mechanism of the exciton tunneling
as a whole entity via the emission of the LO phonon
\cite{RosslerLO1} which should dominate over the separate transfer
of electrons and holes if the tunnel coupling of QWs is strong
(small L$_B$).  Small separation between MW and NMW hh-exciton PLE
bands at high B, realized in the structure with the barrier L$_B$=
3 nm (Fig.~4), allows to record PLE spectra with much better
quality in cross-circular polarizations. Spectral deconvolution by
gaussian bands (Fig.~9(a)) makes possible to analyze tunneling
efficiency between MW and NMW hh-exciton ground states depending
on the exciton energy separation $\Delta E_{ex}$. Figure~9
demonstrates that, starting at B = 0 T, where $\Delta E_{ex} >
\hbar\omega_{LO}$, towards $\Delta E_{ex}=$ 0 no threshold-like
behavior is observed in a relative tunneling efficiency P$_t$,
defined here as a ratio of the MW to NMW hh-exciton PLE band
intensities of the hh-excitons. Oppositely, at $\Delta E_{ex} >
\hbar\omega_{LO}$ value of P$_t$  is even somewhat smaller than
that at $\Delta E_{ex} < \hbar\omega_{LO}$  (Fig.~9(b)). This
result excludes exciton tunneling through the emission of a single
LO phonon, the mechanism theoretically considered by Michl {\it et
al.} \cite{RosslerLO1} for ADQWs with thin barriers and which has
been observed experimentally only in a few experiments
\cite{singleLO,singleLO_B}.

\begin{figure}%h
\begin{center}
\includegraphics[width=8.5cm]{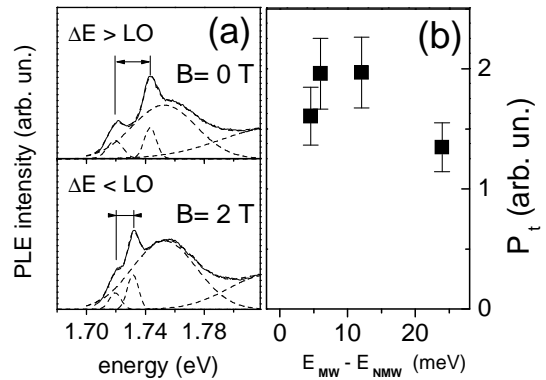}
\end{center}
\caption{ (a) Spectral deconvolution by gaussians of PLE spectra
for ADQW with L$_B$ = 3 nm at B = 0 T ($\Delta E_{ex} >
\hbar\omega_{LO}$) and  B = 2 T ($\Delta E_{ex} <
\hbar\omega_{LO}$). (b) Tunneling probability P$_t$ from the MW to
the NMW QW vs hh-excitons energy separation $\Delta E_{ex}$. }
\label{spectra}
\end{figure}

\subsubsection{Two-step exciton tunneling mechanism}

Previous investigations of tunneling phenomena in ADQWs have shown
that resonant charge transfer in many cases actually involves
exciton tunneling from spatially direct to indirect exciton state
\cite{Ferreira90,Kohler91,Kohler92SPIE, Ferreira90SSc,Ferreira92}.
On the other hand, calculations
\cite{Ferreira90,Henneberger96,RosslerLO1} indicate that tunneling
between spatially direct exciton states is normally not efficient
due to very weak admixture of the direct exciton wave functions.
It has been suggested that two-step tunneling scheme would work
more efficiently with an indirect exciton as an intermediate
state. Experiment has confirmed the effectiveness of the two-step
scheme  when LO phonon emission is allowed at every step thus
overcoming too weak tunneling between direct exciton states
\cite{Henneberger96}.

Calculations show that at low B both indirect excitons have higher
energies than both direct excitons in  ADQWs with L$_B$= 4 and 6
nm  (Fig.~10, L$_B$= 6 nm case), i.e. the above mechanism is
ineffective. The situation changes at B $>$ B$_C^*\approx$ 5 T
when indirect exciton has lower energy. The lowest indirect
exciton is composed of the NMW electron and MW hole, i.e.
symbolically e(NMW)-hh(MW), while another indirect exciton
e(MW)-hh(NMW) is always the highest energy exciton state (Fig.~10)
and thus can be omitted as an intermediate state. First step in
such exciton transfer scheme involves heavy hole tunneling hh(NMW)
$\rightarrow$ hh(MW). The second step involves electron tunneling
e(NMW) $\rightarrow$ e(MW) and it is energetically prohibited in
the single-particle picture for electrons with zero in-plane
kinetic energy  as discussed above. But the two-particle nature of
hole-assisted electron tunneling process makes it different from
the ordinary one-particle (electron or hole) tunneling process
\cite{Yamamoto95}. The prohibition can be overcome by an
additional excitonic Coulombic potential which changes locally
confinement potential of the heterostructure
\cite{Coulombic1,Coulombic2} and may provide electron tunneling to
the final, energy profitable exciton state in the MW. The analysis
of the NM hh-exciton intensity in $\sigma^-$ polarized PLE spectra
(Fig.~3) does not show threshold behavior at B$>$B$_C$, which may
point to the opening of the additional NMW$\rightarrow$MW exciton
transfer channel. Thus, we do not have direct evidence of the
two-step exciton tunneling at high fields.

\begin{figure}%h
\begin{center}
\includegraphics[width=8.5cm]{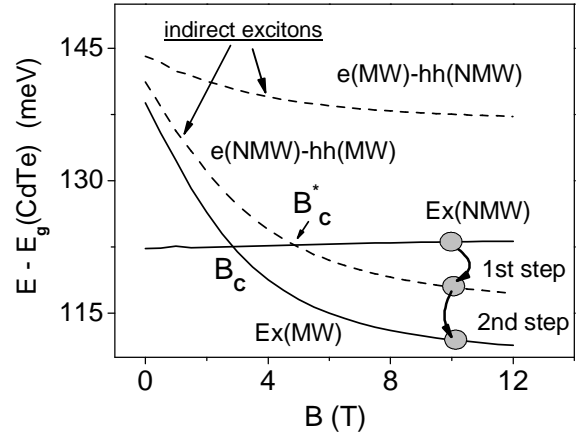}
\end{center}
\caption{ Calculated magnetic field dependencies of direct
(Ex(NMW), Ex(MW)) and indirect (e(MW)-hh(NMW), e(NMW)-hh(MW))
exciton energies in ADQW with L$_B$= 6 nm. Two-step exciton
interwell transfer through the intermediate indirect exciton state
at high B is illustrated schematically. } \label{spectra}
\end{figure}

Comparable intensities of the MW and NMW hh-excitons in PLE
spectra in all studied ADQWs with big range of barrier thickness
points to weak dependency of the tunneling mechanism on the L$_B$.
Interestingly, such an independency has been found in recent
time-resolved investigations in semimagnetic ZnMnSe/ZnCdSe ADQWs
\cite{Toropov}. Very long (few ns) PL decay times observed in Ref.
\cite{Toropov} rule out the possibility of free exciton tunneling
and evidence for the transfer process through some  kind of
localized excitons or excitations. Insensitivity of the decay
times to the barrier height as well as to the barrier width (4--8
nm) contradicts to the traditional exponentially dependent
tunneling mechanism. Authors \cite{Toropov} suggested
photon-exchange energy transfer \cite{Lyo2000} of the localized
excitons or electron-hole pairs. This mechanism is known as slowly
varying with interwell distance \cite{Lyo2000} and that's why is
regarded as the most plausible. Contrary to the PL data of Ref.
\cite{Toropov}, in presented ADQWs exciton recombination from the
MW is absent in PL spectra at B$<$B$_C$ (Fig.~2). This fact
evidences about much quicker exciton transfer in the investigated
ADQWs with thermally induced disorder and points to different
transfer mechanism than that realized in ADQWs studied by Chen
{\it et al.} \cite{Toropov}.

Concluding this discussion, we would like to note that, in
principle, finite uncontrolled doping level in nominally undoped
heterostructures supply carriers which can significantly change
local potential band picture, promoting single-particle exciton
tunneling with the help of carriers localized in the opposite QW.
It is known \cite{defectsCdTe1,defectsCdTe2} that intrinsic
defects in CdTe can provide both n- and p-doping type, with the
doping level  dependent on the growth conditions. High tunneling
efficiency was reported \cite{Kusrayev2} the in CdTe-based ADQWs
with L$_B \leq$ 6 nm and n-type doping concentration  of the order
of 10$^{15}$ cm$^{-3}$. On the other hand, localized carriers
would form different exciton-bound complexes which were present in
the cited work \cite{Kusrayev2}. This is not the case in the
studied structures, which do not show any pronounced PL spectral
bands additional to the localized at potential fluctuations
hh-exciton band (Fig.~2). Thus, we rule out major contribution to
the interwell tunneling due to free/localized carriers supplied by
residual impurities or/and defects.

\subsubsection{Valence band-mixing effects in exciton tunneling}

Stress relaxation during RTA give rise to small values of
$\Delta_{hh-lh}$= 19 and 15 meV in ADQWs with L$_B$= 4 and 6 nm,
respectively (Fig.~4), which are same or smaller than hh-exciton
binding energies $\simeq$ 18 meV (Fig.~7). As a result, valence
band-mixing should be significant in definition of exciton states
\cite{IvchenkoBook,SandersChang}. Theoretical considerations and
numeric calculations have shown that inclusion of valence
band-mixing is crucial for qualitative understanding and
estimation of the hole transfer rates both at on- and
off-resonance conditions \cite{russians,FerreiraELett,Ferreira90}.
Particularly, Ferreira and Bastard \cite{FerreiraELett} have
stressed that the rates of tunneling assisted by elastic
scattering via interface defects, impurities, or alloy
fluctuations, are considerably increased by the valence
band-mixing effect over ones deduced without taking into account
the band-mixing.

An exact solution of the single-particle hh tunneling in ADQW via
acoustic phonons  with accounting of the hh-lh mixing is done in
Ref. \cite{russians}. As it has been shown \cite{russians},
relaxation rate at out-of-resonant conditions is defined by the
light hole tunnel exponent and does not depend on the heavy holes
levels mismatch $\Delta E_h$ when it is bigger than tunnel matrix
element, in  agreement with some experimental data
\cite{Leo90,Kohler91}. It is important that this result is valid
not only for acoustic phonons scattering but also for any
short-range potential scattering relevant in the case of ADQWs
with essential disorder \cite{russians}. Estimations made
according to Ref. \cite{russians} show that in the studied ADQWs
hh tunneling through the lh-mixing channel should prevail at
$\Delta E_h \gtrsim$ 2 meV, i.e. in the whole magnetic field range
except of a narrow field interval near B$_H$, where resonant
tunneling may prevail.

It is a decoupled model that has led to a widespread belief that
hh tunnels considerably slower than electron. Without accounting
of the hh-lh mixing calculated hh transfer times via elastic
scatterers are of several orders longer than the observed ones
\cite{Ferreira90}. hh-lh coupling strongly depends upon the
in-plane wave vector {\em {\bf K}}. Even if the initial state is a
hh {\em {\bf K}}= 0 state, elastic scattering via any static
scatterer can couple hh to a {\em {\bf K}}$\neq$ 0 lh state, which
due to band-mixing has a nonzero projection to the other hh state
\cite{BastardBook}. The importance of band-mixing effect on the
interwell heavy hole tunneling was revealed in experimental
studies in electrically biased ADQWs
\cite{FerreiraELett,Ferreira90SSc}. Numeric calculations, made in
these Refs. \cite{FerreiraELett,Ferreira90SSc}, have shown that
the hole and electron transfer times assisted via the static
elastic scatterers are in general comparable. Moreover, it is
absolutely necessary to account for valence band-mixing for
explanation of resonances in tunneling times between hh and lh
states, observed in Refs. \cite{Kohler91,hhlhRes}.

We suggest that the valence band-mixing effect is a determinant
factor responsible for the observed effective exciton relaxation
in the whole magnetic field range, i.e. at all discussed
experimentally realized band alignment configurations. Spatial
delocalization of the lh states in investigated ADQWs would
promote very efficient interwell tunneling of the hh-excitons
through the strong hh-lh mixing. As for the tunneling mechanism
itself we naturally suppose elastic or quasi-elastic scattering
via static scatterers such as alloy composition fluctuations, in
first order, interface defects and impurities, which are expected
to be present in abundance in the investigated ADQWs as a result
of RTA treatment and ternary composition. Exciton relaxation via
acoustic phonons is of orders weaker (Fig.~8) while LO phonon
assisted exciton tunneling \cite{RosslerLO1} turns out to be
ineffective even in the ADQW with L$_B$= 3 nm (Fig.~9). Full
theoretical description of exciton tunneling via quasi-elastic
scattering with taking into consideration valence band-mixing of
heavy hole is not developed at present. Very interesting attempts
to consider the hh-exciton tunneling problem with accounting of
the exciton Coulomb correlations of the electron-hole pair have
been undertaken only in several publications
\cite{Ferreira90,Ferreira92,Henneberger96,Sugakov} by choosing of
appropriate exciton wavefunctions for initial and final carrir
states but without considering of the valence band-mixing.

Strong influence of valence band-mixing effects on the exciton
tunneling can be suggested from the PL experiments also (Fig.~2).
$\sigma^+$-polarized hh NMW exciton has negligible intensity at
B$>$B$_C$  which evidences about complete transfer of excitons
from this state to the MW during relaxation of photoexcited
carriers. Non-resonantly excited carriers relax in two stages:
after ultrafast subpicosecond energy relaxation by LO-phonon
emission \cite{LOinQW} excitons with a large center-of-mass
momentum {\em {\bf K}} are created during first tens picoseconds
\cite{Damen90}. Then, for hundreds of picoseconds, exciton loses
its excess energy through acoustic phonon emission \cite{Damen90}.
The hole acquires a part of the total exciton momentum and thus
turns up in a strongly valence band mixed state. Short tunneling
times of the hh-exciton as compare to the energy relaxation times
(hundreds of picoseconds) are required to explain the experimental
finding.

\section{CONCLUSIONS}

Magneto-optical properties of semimagnetic Cd(Mg,Mn)Te-based
ADQWs, subjected to the quick thermal annealing, are studied. Very
efficient interwell relaxation of the heavy hole excitons is
observed in the whole magnetic field range regardless of the
tunneling direction. Single-particle description fails to explain
tunneling picture that emphasizes importance of the excitonic
correlations in the tunneling processes in semiconductor
heterostructures.

Main conclusions of the work are the following:

1)  rapid temperature annealing of the ADQW with SiO$_2$ cap layer
effectively introduce magnetic (Mn) and nonmagnetic (Mg) impurity
atoms from the barriers  into initially nonmagnetic CdTe QWs with
a good optical quality. It allows to control mutual carrier levels
alignment and energy order of the intrawell exciton transitions in
such ADQWs by the external magnetic field.

2) Exciton tunneling direction is defined by the exciton
transitions energy difference in the adjacent QWs independently on
the single-particle levels alignment. Thus, excitonic effects are
absolutely important in charge transfer and determine the nature
of tunneling in ADQWs with thermally induced disorder. In turn, it
makes possible to control exciton tunneling direction with
external magnetic field.

3) Absence of anticrossing behaviour of hh-excitons evidences
about incoherent nature of hh-exciton tunneling and the absence of
excitonic level interaction  in ADQWs with strong short-range
disorder, caused by the quick thermal annealing.

4) The simplest, rectangular shape model of ADQW band potential
profile has proved to describe satisfactorily the magnetic field
behavior of the hh- and lh-exciton states.

5) Interwell stress relaxes completely during rapid temperature
annealing in the ADQWs with barriers $\gtrsim$ 4 nm, leading to
small values of hh-lh splitting $\Delta_{hh-lh}$, while in
structures with narrow barriers stress effects remain important.

6) Valence band-mixing effects  in  II-VI semiconductor ADQWs with
strong excitonic effect and small $\Delta_{hh-lh}$ are supposed as
an underlaying physical factor  for the effective exciton
tunneling experimentally observed for different interwell
alignment configurations. Elastic scattering via  alloy
composition fluctuations and interface defects are considered as
the determinant tunneling mechanisms in the ternary ADQWs,
subjected to the rapid temperature treatment.

In general, tunneling experiments in ADQWs, based on II-VI
semimagnetic materials, have shown specificity of the tunneling
phenomena in heterostructures made on these compounds. Strong
excitonic correlations and valence band-mixing are supposed to
govern the nature of interwell exciton transfer. Experimental data
point that the role of valence band-mixing effects in exciton
transfer should be reconsidered for II-VI semiconductor
heterostructures with strong excitonic effect.

\section{ACKNOWLEDGMENTS}

This work was supported by Russian Foundation for Basic Research,
grant No. 04-02-17338.


\begin{thebibliography}{}

\bibitem{Timofeev2004} V.B.~Timofeev, UFN {\bf 174}, 1109 (2004).

\bibitem{Butov1} L.V.~Butov, A.C.~Gossard, and D.S.~Chemla, Nature {\bf 418},
751 (2002).

\bibitem{Butov2} L. V. Butov,
A.L.~Ivanov, A.~Imamoglu, P.B.~Littlewood, A.A.~Shashkin,
V.~T.~Dolgopolov, K.L.~Campman, and A.C.~Gossard, Phys. Rev. Lett.
{\bf 86}, 5608 (2001).

\bibitem{EsakiIEEE} L. Esaki, IEEE J Quantum Electron. {\bf QE-22},
1611 (1986).

\bibitem{Henneberger96} S.~Ten, F.~Henneberger, M.~Rabe,
and N.~Peyghambarian, Phys. Rev. B {\bf 53}, 12637 (1996).

\bibitem{Ferreira92} R. Ferreira, P. Rolland, Ph. Roussignol, C. Delalande, A.
Vinattieri, L. Carraresi, M. Colocci, N. Roy, B. Sermage, and
J.F.~Palmier, B.~Etienne, Phys. Rev. B {\bf 45}, 11782 (1992).

\bibitem{Heimbrodt98}  D.~Suisky, W.~Heimbrodt, C.~Santos, F.~Neugebauer, M.~Happ,
B.~Lunn, J.E.~Nicholls, and D.E.~Ashenford, Phys. Rev. B {\bf 58},
3969 (1998).

\bibitem{DQWth1} I. Galbraith and G. Duggan, Phys. Rev. B {\bf 40}, 5515 (1989).

\bibitem{Henneberger98} W.~Heimbrodt, L.~Gridneva, M.~Happ, N.Hoffmann, M.~Rabe,
and F.~Henneberger, Phys. Rev. B {\bf 58}, 1162 (1998).

\bibitem{Henneberger99} W.~Heimbrodt, M.~Happ, and F.~Henneberger, Phys. Rev. B {\bf 60}, R16326
 (1999).

\bibitem{Heimbrodt2003} H.~Falk, J.~H\"{u}bner, P.J.~Klar, and W.~Heimbrodt, Phys. Rev. B {\bf 68},
165203 (2003).

\bibitem{Ferreira90} R.~Ferreira, C.~Delalande, H.W.~Liu, and G.~Bastard, B.~Etienne, J.F.~Palmier,
Phys. Rev. B {\bf 42}, 9170 (1990).

\bibitem{Oberly89} D.Y.~Oberly, J.~Shah, T.C.~Damen, C.W.~Tu,
T.Y.~Chang, D.A.B.~Miller, J.E.~Henry, R.F.~Kopf, N.~Sauer, and
DiGiovanni, Phys. Rev. B {\bf 40}, 3028 (1989).

\bibitem{Leo90} K.~Leo, J.~Shah, J.P.~Gordon, T.C.~Damen, D.A.B.~Miller, C.W.~Tu,
J.E.~Cunningham, Phys. Rev. B {\bf 42}, 7065 (1990).

\bibitem{Kohler91} M. Nido, M.G.W. Alexander, W.W. R\"{u}hle, and
K. K\"{o}hler, Phys. Rev. B {\bf 43}, 1839 (1991).

\bibitem{Kohler92pps} A.P. Heberle, W.W. R\"{u}hle, and
K. K\"{o}hler, Phys.~Status~Solidi B {\bf 173}, 381 (1992).

\bibitem{JochenQD2001} J. Seufert, M. Obert, G. Bacher, A. Forchel,
T. Passow, K. Leonardi, and D. Hommel, Phys. Rev. B {\bf 64},
121303 (2001).

\bibitem{MarcusQD2003} M. K. Welsch, G. Bacher, H. Schömig, A.
Forchel, S. Zaitsev, C. R. Becker, and L. W. Molenkamp ,
Phys.~Status~Solidi B {\bf 238}, 313 (2003).

\bibitem{ADQDs} R. Heitz, I. Mukhametzhanov, P. Chen, and A.~Madhukar,
Phys. Rev. B {\bf 58}, R10151 (1998).

\bibitem{ADQDs1} Yu.I. Mazur, Zh.M. Wang, G.G. Tarasov, G.J. Salamo, J.W.
Tomm, V. Talalaev, and H. Kissel Phys. Rev. B {\bf 71}, 235313
(2005).

\bibitem{ShahUltrafast} J. Shah, {\it Ultrafast Spectroscopy of
Semiconductors and Semiconductor Nanostructures}, (Springer Series
in Solid-State Sciences, vol. 115, 1999), ch. 7.

\bibitem{Bastard89} R.~Ferreira and G.~Bastard, Phys. Rev. B {\bf 40}, 1074 (1989).

\bibitem{RosslerLO1} F. C. Michl, R. Winkler, and U. R\"{o}ssler,
Solid State Commun. {\bf 99}, 13 (1996).

\bibitem{pss2002} \L.~K$\l$opotowski, M. Nawrocki, S. Ma$\acute{c}$kowski, and E.
Janik, Phys.~Status~Solidi B {\bf 229}, 769 (2003).

\bibitem{russians} F.T.~Vasko and O.E.~Raichev,
Zh. Eksp. Teor. Fiz. {\bf 104}, 3103 (1993) [Sov.Phys. JETF {\bf
77}, 452 (1993)].

\bibitem{singleLO} I.~Lawrence, S.~Haacke, H.~Mariette, W.W.~R\"{u}hle, H.~Ulmer-Tuffigo,
J.~Cibert, and G.~ Feuillet, Phys. Rev. Lett. {\bf 73}, 2131
(1994).

\bibitem{singleLO_B} A.P. Heberle, M. Oestreich, S. Haacke, W.W.
R\"{u}hle, J. C. Maan, and K. K\"{o}hler, Phys. Rev. Lett. {\bf
72}, 1522 (1994).

\bibitem{Kohler92SPIE} A.P. Heberle, W.W. R\"{u}hle, and
K. K\"{o}hler, SPIE Proc: {\bf 1677}, 234 (1992).

\bibitem{Ferreira90SSc} R. Ferreira, H.W. Liu, C. Delalande, J.F. Palmier and B.
Etienne, Surf. Sci. {\bf 229}, 192 (1990).

\bibitem{Kohler91r2} R. Wessel and M. Altarelli, Phys.
Rev. B {\bf 39}, 12802 (1989).

\bibitem{Kohler91r3} E.T. Yu, M.K. Jackson, and T. C. McGill, Appl. Phys. Lett. {\bf 55},
744 (1989).

\bibitem{EsakiAPL} E.E. Mendez, W.I. Wang, B. Ricco and L. Esaki,
Appl. Phys. Lett. {\bf 47}, 415 (1985).

\bibitem{Kohler91r5} M.K. Jackson, M.B. Johnson, D.H. Chow and T.C. McGill, Appl.
Phys. Lett. {\bf 54}, 552 (1989).

\bibitem{Takagahara85} T.~Takagahara, Phys. Rev. B {\bf 31}, 6552 (1985).

\bibitem{Shah96} A.~Tomita, J.~Shah, and R.S.~Knox, Phys. Rev. B {\bf 53}, 10793 (1996).

\bibitem{Lyo2000} S.K.~Lyo, Phys. Rev. B {\bf 62}, 13641 (2000).

\bibitem{Kim96} D.S.~Kim, H.S.~Ko, Y.M.~Kim, S.J.~Rhee,
S.C.~Hohng, Y.H.~Yee, W.S.~Kim, J.S.~Woo, H.J.~Choi, J.~Ihm,
D.H.~Woo, and K.N.~Kang, Phys. Rev. B {\bf 54}, 14580 (1996).

\bibitem{KimSSC} D.S.~Kim, H.S.~Ko, Y.M.~Kim, S.J.~Rhee,
S.C.~Hohng, Y.H.~Yee, W.S.~Kim, J.S.~Woo, H.J.~Choi, J.~Ihm,
D.H.~Woo, and K.N.~Kang, Solid State Commun. {\bf 100}, 231
(1996).

\bibitem{Furdyna} J.K. Furdyna, J.~Appl.~Phys. {\bf 64}, R29 (1988).

\bibitem{RTA1} D.~T\"{o}nnies, G.~Bacher, A.~Forchel, A.~Waag, and G.~Landwehr,
Appl.~Phys.~Lett. {\bf 64}, 5608 (2001).

\bibitem{RTA2} T. Gurung, S. Ma$\acute{c}$kowski, H. E.
Jackson, L. M. Smith, W. Heiss, J. Kossut, and G. Karczewski ,
Phys.~Status~Solidi B {\bf 241}, 652 (2003).

\bibitem{RTAmy} S.~Zaitsev, M.K.~Welsch, H.~Sch\"{o}mig, G.~Bacher, D.V.~Kulakovskii,
A.~Forchel, B. K\"{o}nig, C.R.~Becker, W.~Ossau, and L. W.
Molenkamp, Semicond.~Sci.~Technol. {\bf 16}, 631 (2001).

\bibitem{revDimaYakovlev} D.R.~Yakovlev and K.V.~Kavokin,
Comments~Condens.~Matter~Phys. {\bf 18}, 51 (1996).

\bibitem{Sugakov} S.B~Lev, V.I.~Sugakov, and G.V.~Vertsimakha, J. Phys.: Condens.
Matter 16, 4033 (2004).

\bibitem{IvchenkoBook} E.L. Ivchenko, P.E. Pikus, {\it Superlattices
and Other Heterostructures. Symmetry and Optical Phenomena}
(Springer, Berlin, 1997).

\bibitem{SemenovDyakonov2001} C.Camilleri, F.~Teppe, D.~Scalbert, Y.G.~Semenov,
M.~Nawrocki, M.~Dyakonov, J.~Cibert, S.~Tatarenko, and
T.~Wojtowicz, Phys. Rev. B {\bf 64}, 085331 (2001).

\bibitem{CrookerAwschalom97} S.A.~Crooker, D.D.~Awschalom, J.J.~Baumberg, F.~Flack, and
N.~Samarth,  Phys. Rev. B {\bf 56}, 7574 (1997).

\bibitem{longRlocaliz} U. Jahn, M. Ramsteiner, R. Hey, H. T. Grahn, E.
Runge, and R. Zimmermann, Phys. Rev. B {\bf 56}, R4387 (1997).

\bibitem{bandOffsetWaag} B. Kuhn-Heinrich, W. Ossau, T. Litz, A.
Waag, and G.~Landwehr, J.~Appl.~Phys. {\bf 75}, 8046 (1994).

\bibitem{RTAbandErr} M.K.~Welsch, H.~Sch\"{o}mig, M.~Legge,
G.~Bacher, A.~Forchel, B. K\"{o}nig, C.R.~Becker, W.~Ossau, and L.
W. Molenkamp,  Appl.~Phys.~Lett. {\bf 78}, 2937 (2001).

\bibitem{IFmixin1} J.A. Gaj, W. Grieshaber, C. Bodin-Deshayes, J. Cibert,
G. Feuillet, Y. Merle d'Aubign\'{e}, and A. Wasiela, Phys.~Rev.~B
{\bf 50}, 5512 (1994).

\bibitem{IFmixin2} W. Grieshaber, A. Haury, J. Cibert, Y.
Merle d'Aubign\'{e}, A. Wasiela, and J.A. Gaj, Phys.~Rev.~B {\bf
53}, 4891 (1996).

\bibitem{gFactorDimaYak} A.A. Sirenko, T. Ruf, M. Cardona,
D.R. Yakovlev, W. Ossau, A. Waag, and G. Landwehr, Phys. Rev. B
{\bf 56}, 2114 (1997).

\bibitem{ExRydberg} R.P.~Leavitt and J.W.~Little, Phys.~Rev.~B {\bf 42},
11774 (1990).

\bibitem{BauerAndo88} G.E.W. Bauer and T. Ando, Phys. Rev. B {\bf 38}, 6015 (1988).

\bibitem{MnSegreg} W. Grieshaber, J. Cibert, J.A. Gaj, Y. Merle d'Aubign\'{e} and A. Wasiela,
Phys. Rev. B {\bf 50}, 2011 (1994).

\bibitem{RTAcrystGrowsTonnies} D. T\"{o}nnies, G. Bacher, A. Forchel, A.
Waag, Th. Litz, D. Hommel, Ch. Becker, G. Landwehr, M. Heuken, and
M. Scholl, J. Cryst. Growth {\bf 138}, 362 (1994).

\bibitem{StressReview} S.C. Jain, M. Willander, and H. Maes,
Semicond.~Sci.~Technol. {\bf 11}, 641 (1996).

\bibitem{RTAcrystGrowsKossut} S. Ma$\acute{c}$kowski, Nguyen The Khoi, P. Kossacki, A. Golnik, J. A. Gaj,
A.~Lema$\hat{l}$tre, C. Testelin, C. Rigaux, G. Karczewski, T.
Wojtowicz and J. Kossut, J. Cryst. Growth {\bf 184}, 966 (1998).

\bibitem{tau_semiclass} T.~Tada, A.~Yamaguchi, T.~Ninomiya, H.~Uchiki,
T.~Kobayashi, and T.~Yao , J.~Appl.~Phys. {\bf 63}, 5491 (1988).

\bibitem{tauCdTe} S. Haacke, N.T.~Pelekanos and H.~Mariette, M. Zigone,
A.P.~Heberle, and W.W.~R\"{u}hle, Phys. Rev. B {\bf 47}, 16643
(1993).

\bibitem{tauSpinCdTe} J. Tribollet, F. Bernardot, M. Menant, G. Karczewski, C. Testelin,
and M.~Chamarro, Phys. Rev. B {\bf 68}, 235316 (2003).

\bibitem{phononScatt} W. Wang, A. Zunger, K.A. Mader, Phys. Rev. B {\bf 53},
2010 (1996).

\bibitem{BasuAlloy} P. Ray and P. K. Basu, Phys. Rev. B {\bf 45}, 9169 (1992).

\bibitem{RTAaplQWs} M.K. Welsch, H. Sch\"{o}mig, M. Legge, G. Bacher,
A. Forchel, B. K\"{o}nig, C.R. Becker, W. Ossau, and L.W.
Molenkamp, Appl.~Phys.~Lett. {\bf 78}, 2937 (2001).

\bibitem{RTAapl94Tonnies} D. T\"{o}nnies, G. Bacher, A. Forchel,
A. Waag, and G. Landwehr, Appl.~Phys.~Lett. {\bf 64}, 766 (1994).

\bibitem{MgDiffQWs} A. Arnoult, J. Cibert, S. Tatarenko, and A. Wasiela,
J.~Appl.~Phys. {\bf 87}, 3777 (2000).

\bibitem{Yamamoto95} H. Cao, G. Klimovitch, G. Bj\"{o}rk, and Y. Yamamoto,
Phys. Rev. B {\bf 52}, 12184 (1995).

\bibitem{Coulombic1} J.W. Wu and A.V. Nurmikko,
Phys. Rev. B {\bf 37}, 2711 (1988).

\bibitem{Coulombic2} S. Haacke, N.T. Pelekanos, H. Mariette, M.
Zigone, A.P. Heberle, and W.W. R\"{u}hle, Phys. Rev. B {\bf 47},
16643 (1993).

\bibitem{Toropov} W.M. Chen, I.A. Buyanova, K. Kayanuma, K. Nishibayashi,
K. Seo, A. Murayama, Y. Oka, A.A. Toropov, A.V. Lebedev, S.V.
Sorokin, and S.V.~Ivanov, Phys. Rev. B {\bf 72}, 073206 (2005).

\bibitem{defectsCdTe1} M.A.~Berding, Phys. Rev. B {\bf 60}, 8943 (1999).

\bibitem{defectsCdTe2} N.~Krsmanovic, K.G.~Lynn, M.H.~Weber, R.~Tjossem,
Th.~Gessmann, Cs.~Szeles, E.E.~Eissler, J.P.~Flint, and
H.L.~Glass, Phys. Rev. B {\bf 62}, R16279 (2000).

\bibitem{Kusrayev2} Yu.G. Kusrayev, A.V. Koudinov, B.P. Zakharchenya, W.E.
Hagston, D.E. Ashenford, B. Lunn., Solid State Commun. {\bf 95},
149 (1995).

\bibitem{SandersChang} G.D. Sanders and Y.C. Chang, Phys. Rev. B {\bf 31}, 6892 (1985).

\bibitem{FerreiraELett} R.~Ferreira and G.~Bastard, Europhys. Lett. {\bf 10}, 279 (1989).

\bibitem{BastardBook} G. Bastard, J. A. Brum, and R. Ferreira, in
 {\it Solid State Physics: Advances in Research and Applications}, edited
by H. Ehrenreich and D. Turnbull (Academic, New York, 1991), Vol.
44, ch. 12.

\bibitem{hhlhRes} T.B. Norris, N. Vodjdani, B. Vinter, E.
Costard, and E. B\"{o}ckenhoff, Phys. Rev. B {\bf 43}, 1867
(1991).

\bibitem{LOinQW} B.P.~Zakharchenya, P.C.~Kop'ev, D.N.~Mirlin, D.G.~Polacov, I.I.~Reshina,
V.F.~Sapega, and A.A.Sirenko, Solid State Commun. {\bf 69}, 203
(1989).

\bibitem{Damen90} T.C. Damen, J. Shah, D.Y. Oberli, D.S.
Chemla, J.E. Cunningham, and J.M. Kuo, Phys. Rev. B {\bf 42}, 7434
(1990).

\end{thebibliography}
\end{document}